\documentstyle[aps]{revtex}

\title{\bf Gauge Independence of Limiting Cases of \\
One-Loop Electron Dispersion Relation in High-Temperature QED}

\author{Indrajit Mitra\thanks{E-mail: indra@tnp.saha.ernet.in}\\ 
\sl{Theory Group, Saha Institute of Nuclear Physics, 1/AF Bidhannagar,\\
Calcutta 700064, India}}

\begin{document}
\maketitle
\def\be{\begin{equation}}
\def\ee{\end{equation}}
\def\bea{\begin{eqnarray}}
\def\eea{\end{eqnarray}}
\def\lb{\label}

%%%%%%%%%%%%%%%%%%%%%%%%%%%%%%%%%%%%%%%%%%%%%%%%%%%%%%%%%%%%%%%%%%%%%%%%%% 

\begin{abstract}
Assuming high temperature and taking subleading temperature dependence
 into account, gauge dependence of one-loop electron dispersion relation is
investigated in massless QED at zero chemical potential. The analysis is carried
out using a general linear covariant gauge. The equation
governing the gauge dependence of the dispersion relation is obtained and used to 
prove that
the dispersion relation is gauge independent in the limiting case of
momenta much larger than $eT$. It is also shown that the 
effective mass is not influenced by the leading temperature dependence of
 the gauge dependent part of the effective self-energy. As a result
 the effective mass, which is of order $eT$, does not receive
a correction of order $e^2T$ from one loop, independent of the gauge parameter.
\end{abstract}

%%%%%%%%%%%%%%%%%%%%%%%%%%%%%%%%%%%%%%%%%%%%%%%%%%%%%%%%%%%%%%%%%%%%%%%

\section {Introduction}  \lb{sec:intro}

It is well-known that the dispersion relation of relativistic electrons in
 material medium is markedly different from that in vacuum, a fact which makes
the study of the former interesting and important. The required calculation is
performed within the framework of statistical quantum field theory (SQFT), also
known as finite-temperature field theory. In order to deduce the dispersion
 relation analytically, it is useful to assume the temperature $T$ to be high. It
 is then found that while the part of the one-loop effective 
 self-energy of the electron (the effective self-energy being the field-theoretic input to 
the dispersion relation; see Sec. \ref{sec:dispersion}) leading in $T$ 
goes like $T^2$ and is gauge independent, the gauge 
dependent part goes like $T$.
This shows that the one-loop dispersion relation is gauge independent to leading
 order in $T$ \cite{weldon} .

A somewhat similar situation occurs for the one-loop effective self-energy of
 neutrinos in material medium, which is gauge independent at $O(1/M{_W^2})$ but gauge
dependent when the $O(1/M{_W^4})$ terms are included. However, it has been shown
 \cite{nieves} that the one-loop dispersion relation following from the latter is still
independent of the gauge parameter.
This raises the possibility that the gauge dependence of the part of the
 one-loop effective self-energy of the electron subleading in $T$ may also not show up in the one-loop
 dispersion relation. This possibility is explored in the present paper in two
different limiting cases. We show that this indeed happens in the $k\gg eT$
 limit of the dispersion relation. In the zero-momentum limit we find that
the gauge dependence of the part of the effective self-energy going like $T$ does not
 contribute to the effective mass. This leads to the gauge independent result
 that the effective mass,
known to be of $O(eT)$ to leading order in $T$ \cite{weldon}, receives no $O(e^2T)$
 correction from one loop.

We consider the electrons to be massless in vacuum (which is true in the phase of restored
 chiral symmetry, and a good approximation when $T$ is much larger than the electron mass), and the chemical
potential of the background medium to be zero. Our calculation makes use of the real-time
 formulation of SQFT.

The demonstration of gauge independence of the properties of elementary particles
in material medium has turned out to be of extreme importance in SQFT. In particular,
the gauge dependence of the gluon damping rate at one loop was a long-standing 
problem. In Ref. \cite{kkr}, the Ward identities determining the gauge dependence
of the gluon dispersion relations were deduced and used to prove gauge independence
in a self-consistent perturbative expansion. In Refs.
\cite{bp1} and \cite{ft}, it was shown that such a consistent expansion requires
resummation of the hard thermal loops. The use of this idea finally
led to a gauge-independent gluon damping rate to leading order in the coupling
constant \cite{bp2}. It will be seen that the
gauge dependence identities of Ref. \cite{kkr}, although dealing with gluons,
are still relevant for the case studied by us.

The plan of the paper is as follows. In Sec. \ref{sec:dispersion}, the general form of the
 fermionic
 dispersion relation is discussed. In Sec. \ref{sec:selfenergy}, we arrive at the expression for
 the one-loop effective electron self-energy in a general linear covariant gauge. 
In Sec. \ref{sec:governingeqn}, the gauge dependence equation for the one-loop dispersion
relation is obtained and the possibility of a generalization of it to all orders
is mentioned.
Gauge independence of the $k\gg eT$ limit of the one-loop dispersion relation
is proved in Sec. \ref{sec:proof-for-high}, while gauge independence of the one-loop effective mass 
is discussed in Sec. \ref{sec:proof-for-mass}. We present our conclusions in Sec. \ref{sec:conclusions}. 
Appendices \ref{app:calculation} and  \ref{app:masseqn} contain some details of the
calculations,  Appendices \ref{app:a-and-b} and \ref{app:corrections} clarify certain points,
 and Appendix \ref{app:massformula}
contains a derivation of the formula for the
effective mass.

\section {General Form of Fermionic Dispersion Relation}    \lb{sec:dispersion}

In the real-time formulation of SQFT the free-field propagator, the full or 
exact propagator and the self-energy are all $2\times 2$ matrices \cite{landsman}. The
poles of the full propagator for a massless fermion of four-momentum $K^\mu$ 
occur at the poles of the function 
\be
S(K)=[\rlap/K -\Sigma(K)]^{-1}
\ee
where $\Sigma(K)$ is related to the components of the self-energy matrix.
Considering the analogy to the vacuum case, we shall call $\Sigma(K)$ the
{\it effective self-energy} of the fermion. It has the general form \cite{weldon}
\be
\Sigma(K)=-a\rlap/K -b\rlap/u                                               \lb{eq:1}
\ee
in a medium with four-velocity $u^\mu$. Here $a$ and $b$ are Lorentz-invariant functions of the
 two Lorentz scalars
\be
\omega\equiv K\cdot u,
\ee
\be
k\equiv [(K.u)^2-K^2]^{1/2}
\ee
$\omega$ and $k$ are the Lorentz-invariant energy and three-momentum
 respectively, which satisfy
\be
K^2=\omega^2-k^2.                                       \lb{eq:K^2}
\ee
The pole in $S(K)$ is then given by
\be
f(\omega,k)\equiv(\omega-k)(1+a)+b=0.                                 \lb{eq:27a}
\ee
In proceeding from this equation, one writes \cite{damping}
\be
\omega=\omega_R - i\frac{\gamma}{2}
\ee
with real $\omega_R$ and $\gamma$, where $\omega_R$ is the (real) fermion
energy and $\gamma$ the fermion damping rate. Equation (\ref{eq:27a}) is thus a relation
between this complex $\omega$ and (real) $k$; the relation between $\omega_R$
and $k$ is the fermionic dispersion relation.

Following Ref.\cite{damping} let us now make the decomposition
\be
a(\omega,k)=a_R(\omega,k)+i a_I(\omega,k)     \lb{eq:5555}
\ee
and likewise for $b$ and $f$. Thus
\be
f_R(\omega,k)=(\omega-k)(1+a_R)+b_R,                          \lb{eq:freal}
\ee 
\be
f_I(\omega,k)=(\omega-k)a_I+b_I.                              \lb{eq:fim}
\ee
Also decompose the function $\Sigma$ as
\be 
\Sigma={\rm Re}\Sigma +i {\rm Im}\Sigma,
\ee
\be
{\rm Re}\Sigma=-a_R\rlap/K - b_R\rlap/u,               \lb{eq:4444}
\ee
\be
{\rm Im}\Sigma=-a_I\rlap/K - b_I\rlap/u.
\ee
We now consider
\be
f_R(\omega_R-i\frac{\gamma}{2},k) + i f_I(\omega_R-i\frac{\gamma}{2},k)=0.
                                                \lb{eq:X3}
\ee
The damping rate $\gamma$ is absent at the tree level and arises from loops. So
it is useful to perform Taylor expansion of the LHS of (\ref{eq:X3}) in
$\gamma$ about $\omega=\omega_R$, as the
powers of $\gamma$ higher than what is required can be neglected. Keeping upto
terms linear in $\gamma$ and $f_I$ ($f_I$ also arising only from loops), we arrive at
\cite{damping} the following two real relations from the complex 
equation (\ref{eq:X3}):
\be
f_R(\omega_R,k)=0                       \lb{eq:Y}
\ee
which is the dispersion ($\omega_R$-$k$) relation to this order, and
\be
\frac{\gamma}{2}\Bigg[\frac{\partial f_R}{\partial \omega}\Bigg]_
{\omega=\omega_R} = f_I(\omega_R,k)               \lb{eq:Z}
\ee
which (on using (\ref{eq:Y})) gives $\gamma$ as a function of $k$ to the same
 order. When the next higher order correction is incorporated into
 (\ref{eq:Y}) from (\ref{eq:X3}), we obtain
\be
f_R(\omega_R,k) +
\frac{\gamma}{2}\Bigg[\frac{\partial f_I}{\partial \omega}\Bigg]_
{\omega=\omega_R}
-\frac{\gamma^2}{8}\Bigg[\frac{\partial^2 f_R}{\partial \omega^2}\Bigg]_
{\omega=\omega_R} = 0.                        \lb{eq:ZZ}
\ee
Since $f_I$ (see (\ref{eq:fim})) and hence $\gamma$ (see (\ref{eq:Z})) are
both of order $e^2$, the last two terms on the LHS of (\ref{eq:ZZ})
are of order $e^4$. The {\it one-loop} 
dispersion relation is thus given
by (\ref{eq:Y}) and (\ref{eq:freal}). To study it {\it we need to consider
only the $Re\Sigma$ part of $\Sigma$}. Therefore (with the exception of
Appendix \ref{app:corrections}) {\it the subscripts in
$a_R$, $b_R$, $f_R$ and $\omega_R$ will be dropped henceforth
and will be assumed to be understood.}

\section {One-Loop Effective Electron Self-Energy in a General Linear 
Covariant Gauge}
                    \lb{sec:selfenergy}

${\rm Re}\Sigma$ is related to the $2\times 2$ self-energy matrix by \cite{landsman}
\be
{\rm Re}\Sigma={\rm Re}\Sigma_{11}.                      \lb{eq:Sigma11}
\ee
$\Sigma_{11}$ is obtained from the standard QED Feynman diagram for the electron
self-energy:
\be
\Sigma_{11}(K)=ie^2\int \frac{{d^4}p}{(2\pi)^4}D_{\mu\nu}(p)\gamma^{\mu}S(p+K)
\gamma^\nu                                      \lb{eq:7}
\ee
where the physical (``11") free-particle massless fermion propagator at zero 
chemical potential is given by 
\be
S(p)=\rlap/p\Bigg[\frac{1}{p^2 + i\epsilon} +2\pi i \delta(p^2) f_F(p)\Bigg],
                          \lb{eq:8}
\ee
\be
f_F(p)=[e^{|p.u|/T} +1]^{-1}                                \lb{eq:9}
\ee
and the physical (``11") free photon propagator in a general covariant gauge 
is given by
\cite{kowalski}
\be
D^{\mu\nu}(p)= D{^{\mu\nu}_{\rm{FG}}}(p)+ D{^{\mu\nu}_\xi}(p),
\ee
\be
D{^{\mu\nu}_{\rm{FG}}}(p)=-g^{\mu\nu}\Bigg[\frac{1}{p^2 + i\epsilon} -2\pi i \delta(p^2) f_B(p)\Bigg],
                                 \lb{eq:11}
\ee
\be
 D{^{\mu\nu}_\xi}(p)=-\xi p^\mu p^\nu \Bigg[\frac{1}{(p^2 + i\epsilon)^2}
       +2\pi i f_B(p)\frac{ d\delta(p^2)}{dp^2}\Bigg],           \lb{eq:12}
\ee
\be
f_B(p)=[e^{|p.u|/T} -1]^{-1}.                                  \lb{eq:13}
\ee
Here $\xi$ is the gauge parameter and FG denotes the Feynman gauge ($\xi=0$).
Since the calculation will be performed in the rest-frame of the medium, in
the rest of this paper we shall 
set $p\cdot u=p_0$ in (\ref{eq:9}) and (\ref{eq:13}).

Let us write
\be
\Sigma=\Sigma^{T=0} + \Sigma^\prime.                         \lb{eq:sigmas}
\ee
We now evaluate the real parts of these functions (real in the sense as 
defined through (\ref{eq:4444}) and (\ref{eq:5555})) by putting the expressions 
for the fermion
and the photon propagators in (\ref{eq:7}) and using (\ref{eq:Sigma11}). 

First consider the $T=0$ part:
\be
{\rm Re}\Sigma^{T=0}=-a^{T=0}(K^2)\rlap/K.                \lb{eq:sigmazero}
\ee
The infinity in $a^{T=0}$ (recall that this is to be understood to denote just the
real part) is removed by the counter-term of
wave-function renormalization, leading to
\be
a{^{T=0}_{ren}}(K^2)=-(1+\xi)\frac{e^2}{(4\pi)^2}\ln\frac
                               {K^2}{\sigma^2}.      \lb{eq:azero}
\ee
where $\sigma$ is the renormalization scale. 
Some details of this are given in Appendix \ref{app:calculation}.

We then turn to the $T$-dependent part, and write
\be
\Sigma^\prime=\Sigma{^\prime_{\rm{FG}}} + \Sigma{^\prime_\xi}.        
\ee
To deal with ${\rm Re}\Sigma
{^\prime_\xi}$, we make use of
\be
\frac{d\delta(p^2)}{dp^2}=\frac{1}{2\pi i}\Bigg[\frac{1}{(p^2+i\epsilon)^2}
             -\frac{1}{(p^2-i\epsilon)^2}\Bigg]
\ee
and then use the regularization 
\be
\frac{1}{(p^2\pm i\epsilon)^2}=\lim_{\lambda\rightarrow 0}\frac{\partial}
{\partial\lambda^2}\frac{1}{p^2-\lambda^2\pm i\epsilon}     \lb{eq:reg}
\ee
so that
\be
\frac{d\delta(p^2)}{dp^2}=-\lim_{\lambda\rightarrow 0}\frac{\partial}
{\partial\lambda^2}\delta(p^2-\lambda^2).              \lb{eq:delta}
\ee
Following the manipulations described in Appendix \ref{app:calculation},
we arrive at 
\be
{\rm Re}\Sigma{^\prime_{\rm FG}}= 2e^2\int\frac{d^4p}{(2\pi)^4}
     [(\rlap/p+\rlap/K)f_B(p) + \rlap/p f_F(p)] 
     2\pi\delta(p^2){\rm Re}\frac{1}{2p.K+K^2+i\epsilon},    \lb{eq:16}
\ee
\be
{\rm Re}\Sigma{^\prime_\xi} = {\rm Re}\Sigma{^\prime_{\xi,F}} +
            {\rm Re}\Sigma{^\prime_{\xi,B}},                 \lb{eq:17}
\ee
\be
{\rm Re}\Sigma{^\prime_{\xi,F}} = \xi e^2\int\frac{d^4 p}{(2\pi)^4}
    [K^2 \rlap/p-2p.K\rlap/K] 2\pi\delta(p^2)f_F(p)
    {\rm Re}\frac{1}{(2p.K + K^2 +i\epsilon)^2}, 
                                  \lb{eq:21}
\ee
\bea
{\rm Re}\Sigma{^\prime_{\xi,B}} = \xi \frac{e^2}{2}\int\frac{d^3 \vec{p}}
       {(2\pi)^3}\frac{1}{e^{\beta |\vec{p}|}-1}\frac{1}{|\vec{p}|}
      \Bigg[\Bigg(\rlap/K+\frac{K^2\vec{p}.\vec{\gamma}}{2|\vec{p}|^2}\Bigg)
      {\rm Re}\Bigg(\frac{1}{D_+}+\frac{1}{D_-}\Bigg)    \nonumber  \\
      -K^2 |\vec{p}|{\gamma}_0 {\rm Re}\Bigg(\frac{1+K_0/|\vec{p}|} {{D_+}^2}
      -\frac{1-K_0/|\vec{p}|}{{D_-}^2}\Bigg)    \nonumber  \\
      +K^2 \vec{p}.\vec{\gamma} {\rm Re}\Bigg(\frac{1+K_0/|\vec{p}|}{{D_+}^2}
      +\frac{1-K_0/|\vec{p}|}{{D_-}^2}\Bigg)\Bigg]    \nonumber   \\
      - \xi \frac{e^2\beta}{4}\int\frac{d^3 \vec{p}} {(2\pi)^3}
      \frac{e^{\beta |\vec{p}|}}{(e^{\beta |\vec{p}|}-1)^2}\frac{1}{|\vec{p}|^2}
      \Bigg[K^2|\vec{p}|\gamma_0 {\rm Re}\Bigg(\frac{1}{D_+}-\frac{1}{D_-}\Bigg
       )          \nonumber  \\
      -K^2\vec{p}.\vec{\gamma}{\rm Re}\Bigg(\frac{1}{D_+}+\frac{1}{D_-}\Bigg)
      \Bigg],                           \lb{eq:23}
\eea
where, in the last equation,
\be
D{_\pm} = K^2\pm 2|\vec{p}|K_0 - 2\vec p.\vec K + i\epsilon.
\ee
While the expressions (\ref{eq:16}) and (\ref{eq:21}) are the same as those in
Ref. \cite{weldon} (except that we have carefully incorporated the $i\epsilon$,
 which will be needed for the calculations in Appendix \ref{app:masseqn}), the
expression (\ref{eq:23}) is a different one. We have found this form of
${\rm Re}\Sigma{^\prime_{\xi,B}}$ convenient for actually carrying out the 
integration (this, again, is done in Appendix \ref{app:masseqn}).

Let us now investigate the high-$T$ behaviour of (\ref{eq:16}), (\ref{eq:21})
and (\ref{eq:23}). Following Ref. \cite{weldon}, this behaviour can be inferred
from the degree of UV (ultraviolet) divergence of the integral in each
expression in the absence of the cut-off of $O(T)$ provided by $f_B$ or $f_F$
or $e^{\beta|\vec p|}/(e^{\beta|\vec p|} - 1)^2$. One then finds that
${\rm Re}\Sigma{^\prime_{FG}}$ goes like $T^2$, while
${\rm Re}\Sigma{^\prime_{\xi,F}}$ and ${\rm Re}\Sigma{^\prime_{\xi,B}}$ 
go like $T$. It may be noted that one cannot obtain the correct high-$T$
behaviour of ${\rm Re}\Sigma{^\prime_{\xi,B}}$ without removing the regulator
$\lambda$, and this can be done only after performing the $p_0$-integration,
as in (\ref{eq:23}).

At high $T$, one can neglect the renormalized ${\rm Re}\Sigma^{T=0}$
compared to ${\rm Re}\Sigma^\prime$ \cite{weldon1}. Actually ${\rm Re}\Sigma^{T=0}$
depends on the renormalization scale $\sigma$, but since the dependence is 
logarithmic, we can still ignore it vis-a-vis the power law dependences on $T$
in the various parts of ${\rm Re}\Sigma^\prime$. We shall, however, see that
it is {\it not necessary} to neglect ${\rm Re}\Sigma^{T=0}$ for proving
gauge independence at $k\gg eT$. We shall also use (\ref{eq:azero}) to
incorporate the $\sigma$-dependence in the equation for the effective mass.

\section {Equation Governing Gauge Dependence of Dispersion Relation at One Loop,
          and General Considerations}
                 \lb{sec:governingeqn}

Inverting (\ref{eq:1}), one obtains (recall that we are considering only the
 ${\rm Re}\Sigma$ part) 
\be
a=\frac{1}{4k^2}[{\rm Tr}(\rlap/K{\rm Re}\Sigma)-
         \omega{\rm Tr}(\rlap/u{\rm Re}\Sigma)],  \lb{eq:24}
\ee
\be
b=\frac{1}{4k^2}[-\omega {\rm Tr}(\rlap/K{\rm Re}\Sigma)+
        (\omega^2-k^2) {\rm Tr}(\rlap/u{\rm Re}\Sigma)].        \lb{eq:25}
\ee
The dispersion relation is obtained by putting $a$,$b$ in (\ref{eq:27a}).
Let us now write the function $f$ in (\ref{eq:27a}) as
\be
f=f_{\rm{FG}}+f_{\xi},
\ee
\be
f_{\rm{FG}}=(\omega-k)(1+a_{\rm{FG}})+b_{\rm{FG}},
\ee
\be
f_{\xi}=(\omega-k)a_{\xi}+b_{\xi}.                      \lb{eq:29}
\ee
Next using (\ref{eq:17}), (\ref{eq:21}) and (\ref{eq:23}), one readily sees
 that $K^2$ factors out from the expression for
${\rm Tr}(\rlap/K {\rm Re}\Sigma{^\prime_\xi})$. 
Therefore in view of (\ref{eq:K^2}), one can write from (\ref{eq:25}) that
\be
b_\xi=(\omega-k)(\mbox{1-loop function of $\omega$,$k$})
\ee
(note that $b$, being zero at $T=0$, is determined by $\Sigma^\prime$).
Consequently
\be
f_\xi= (\omega-k)[a_\xi + (\mbox{1-loop function of $\omega$,$k$})]
           \lb{eq:gov}.
\ee
This is the equation governing the gauge-dependence of the electron dispersion 
relation at one-loop. This equation is of the form
\be
(f_\xi)_{\mbox{1-loop}}=f_{\mbox{tree}}\times\mbox{[1-loop $\xi$-dependent function]}.
                        \lb{eq:gov1}
\ee
As we shall see in the next section, {\it the fact that $f_{\rm tree}$ 
factors out on the RHS of Eq. (45) is
crucial to the proof of gauge independence.}

Ref. \cite{kkr} contains a general, nonperturbative derivation of the identities
determining the gauge dependence of the gluon dispersion relations (see Eqs. (16) 
and (17) of \cite{kkr}). To one loop these gauge dependence identities are shown to
reduce to relations (see Eq. (20) of \cite{kkr}) which are analogous to Eq.
(\ref{eq:gov1}) above. {\it It is therefore likely that a general
gauge dependence identity for the electron dispersion relation, 
reducing to Eq.} (45) {\it at one loop, can be derived and used to
arrive at a general proof of gauge independence of the dispersion relation}
(as in the gluon case).
In this paper,
however, we confine ourselves to one-loop calculations.

\section {Gauge Independence of One-Loop Dispersion Relation at Momenta Much 
Larger than $\lowercase{e}T$}                 \lb{sec:proof-for-high}

Since we shall consider $k\gg eT$ in this section, let us first note that
for {\it sufficiently small e} there always exists a
domain $eT\ll k\ll T$, so that the restriction $k\gg eT$ does not 
contradict the
high-$T$ approximation.
 
Retaining only the terms with leading $T$-dependence in $a$ and $b$,the following
one-loop dispersion relation was obtained in Ref. \cite{weldon} for $k\gg eT$:
\be
\omega=k+\frac{{M_0}^2}{k}-\frac{{M_0}^4}{2k^3}\ln\frac{2k^2}
                 {{M_0}^2}+ \cdots   \lb{eq:27}
\ee
where ${M_0}^2=e^2T^2/8$. The terms on the RHS clearly indicate that the 
above expansion is valid for $k\gg M_0$ i.e. $k\gg eT$.
As shown in Ref.\cite{weldon} and as we shall see in Sec.\ref{sec:proof-for-mass}, 
$M_0$ is the leading-order effective
 electron mass \cite{symbol} and thus can be considered a measure of 
finite-temperature effects. With $k\gg eT$ ensuring that the finite-temperature 
{\it effects}
are small (even though $T$ is large), the equation (\ref{eq:27a}) in the
present case
has resulted in a deviation from $\omega=k$ by powers of $e^2$ and of 
$e^2({\rm ln}(1/e))$ \cite{footnote1}. 
This can be understood from (\ref{eq:27a}) in view of the fact 
that $a$ and $b$ are in this case primarily of $O(e^2)$ which may come multiplied with
powers of
${\rm ln}(1/e)$ (originating from powers of ${\rm ln}(\omega-k)$).

The expected general form of the one-loop dispersion relation for $k\gg eT$
{\it when the terms subleading in $T$ are kept} is
\be
\omega=k+e^2f_1(e,k,T)                        \lb{eq:26}
\ee
{\it where the $e$-dependence of $f_1$ involves only powers of $\:\ln(1/e)$}.
Equation (\ref{eq:27}) (without the last term on its RHS, which is actually of 
order $e^4 \ln(1/e)$) is a special case of (\ref{eq:26}), 
where $f_1$ turned out
to be totally $e$-independent. Let us now take (\ref{eq:26}) to be the
relation in the Feynman gauge i.e. assume that it satisfies $f_{\rm FG}=0$.
Then to prove that (\ref{eq:26}) is gauge independent we have to show that
it also satisfies that $f_{FG}+f_\xi=0$. It suffices to show that, at
(\ref{eq:26}), $f_\xi$ is of order $e^4$ (with or without
powers of $\ln(1/e)$) \cite{similar}. 
This follows readily
from (\ref{eq:gov}), since the portion of the RHS of (\ref{eq:gov})
within the square brackets involves terms of the order of $e^2$ and 
$e^2({\rm ln}(1/e))^n$ ($n$ is an integer), and so also does 
$(\omega-k)$ at (\ref{eq:26}).

Digressing briefly from {\it high-temperature} QED, we mention that {\it the 
above demonstration of gauge independence is actually sufficient also for the 
case which involves} not large $T$ but {\it large momenta}. Thus for $k\gg T$,
a one-loop relation of the key form (\ref{eq:26}) should still hold. Note that
the rest of the proof, namely, arriving at (\ref{eq:gov}), did not assume large 
$T$ (in particular, we did not neglect the vacuum contribution to $a$). The
proof of gauge independence continues to be valid if, in addition, we have a
chemical potential $\mu$ such that $k\gg \mu$ as well. Then the relevant changes are 
that $f_1$ in (\ref{eq:26})
depends on $\mu$ as well, $f_F$ is modified and we have $f_F(-p)$ in place of
$f_F(p)$ in (\ref{eq:21}), but none of these affect the proof outlined above.

\section {Gauge Independence of Effective Mass at One Loop}   \lb{sec:proof-for-mass}

The effective electron mass is the value of $\omega$ at $k=0$. The analysis of 
the previous section does not prove the gauge independence of the effective mass
because the form (\ref{eq:26}) does not hold near $k=0$. For example, to leading order
in $T$ the one-loop dispersion relation for $k\ll eT$ is \cite{weldon}
\be
\omega=M_0+\frac{k}{3}+\frac{k^2}{3M_0}+\cdots
\ee
This suggests that {\it the leading order values of $a$ and $b$ are not of $O(e^2)$}
(with or without powers of ${\rm ln}(1/e)$) {\it in this limit}; indeed the values
at $k=0$ are $a=-1/3$ and $b=-2M_0/3$ \cite{footnote2}. In Appendix
\ref{app:a-and-b} we show how this apparently surprising behaviour can be
understood in a simple way.

To investigate the gauge dependence of the effective mass $M$ we shall use
the following relation \cite{erdas}:
\be
M^2=\frac{1}{4}\lim_{k\rightarrow 0,\omega\rightarrow M}
           {\rm Tr}[\rlap/K{\rm Re}\Sigma].      \lb{eq:31}
\ee
In the rest frame of the medium, the above limit translates to putting
$\vec{K}=\vec{0}$ and $K_0=M$. A careful derivation of (\ref{eq:31}) in the
present context is provided
in Appendix \ref{app:massformula}.

On using (\ref{eq:sigmas}) and (\ref{eq:sigmazero}) in (\ref{eq:31}), we obtain
\be
M^2=-M^2a^{T=0}(M^2)+{M^\prime}^2,             \lb{eq:totalmass}      
\ee
\be
{M^\prime}^2\equiv\frac{1}{4}\lim_{k\rightarrow 0,\omega\rightarrow M}
           {\rm Tr}[\rlap/K{\rm Re}\Sigma^\prime].      \lb{eq:Tmass}
\ee
At $T=0$, $M^\prime=0$ and so (\ref{eq:totalmass}) is correctly satisfied by 
$M=0$.

Putting in (\ref{eq:Tmass}) the expressions for ${\rm Re}\Sigma{^\prime_{\rm{FG}}}$,
${\rm Re}\Sigma^\prime_{\xi,F}$ and ${\rm Re}\Sigma^\prime_{\xi,B}$ given by (\ref{eq:16}),
(\ref{eq:21}) and (\ref{eq:23}), we get
\be
{M^\prime}^2=M{^\prime_{\rm{FG}}}^2+M{^\prime_{\xi,F}}^2+
             M{^\prime_{\xi,B}}^2,         \lb{eq:three}
\ee
\be
M{^\prime_{\rm{FG}}}^2=2e^2\int\frac{d^4p}{(2\pi)^3}[(p_0+M)f_B(p)+p_0f_F(p)]
               \delta(p^2){\rm Re}\frac{1}{2p_0+M+i\frac{\epsilon}{M}},
                                         \lb{eq:33}
\ee
\be
M{^\prime_{\xi,F}}^2=-\xi e^2 M\int\frac{d^4p}{(2\pi)^3}p_0 f_F(p)
               \delta(p^2){\rm Re}\frac{1}{(2p_0+M+i\frac{\epsilon}{M})^2},
                                         \lb{eq:34}
\ee
\be
M{^\prime_{\xi,B}}^2= M{^\prime_{\xi,B({\rm I})}}^2+ M{^\prime_{\xi,B({\rm II})}}^2+ 
                      M{^\prime_{\xi,B({\rm III})}}^2,
                                       \lb{eq:mb}
\ee
\be
M{^\prime_{\xi,B({\rm I})}}^2=\frac{\xi e^2 M}{2}\int\frac{d^3\vec p}{(2\pi)^3}
                \frac{1}{|\vec p|}\frac{1}{e^{\beta|\vec p|}-1}{\rm Re}
                \Bigg(\frac{1}{E_+}+\frac{1}{E_-}\Bigg),   \lb{eq:mb1}
\ee
\be
M{^\prime_{\xi,B({\rm II})}}^2=-\frac{\xi e^2 M}{2}\int\frac{d^3\vec p}{(2\pi)^3}
                \frac{1}{e^{\beta|\vec p|}-1}{\rm Re}
                \Bigg[\frac{M}{|\vec p|}\Bigg(\frac{1}{{E_+}^2}+\frac{1}{{E_-}^2}\Bigg)
                +\Bigg(\frac{1}{{E_+}^2}-\frac{1}{{E_-}^2}\Bigg)\Bigg]
                                                         \lb{eq:mb2}
\ee
\be
M{^\prime_{\xi,B({\rm III})}}^2=-\frac{\xi e^2 M^2\beta}{4}\int\frac{d^3\vec p}{(2\pi)^3}
                \frac{1}{|\vec p|}\frac{e^{\beta|\vec p|}}{(e^{\beta|\vec p|}-1)^2}{\rm Re}
                \Bigg(\frac{1}{E_+}-\frac{1}{E_-}\Bigg).
                                                \lb{eq:mb3}
\ee
Here
\be
E_\pm=M\pm2|\vec p|+\frac{i\epsilon}{M}.
\ee

We now investigate the high-$T$ behaviour of the three terms on the RHS of 
(\ref{eq:three}). This can be arrived at from the UV behaviour, as explained
towards the end of Sec. \ref{sec:selfenergy}. However there is a 
constraint that each term on the RHS of (\ref{eq:three}) is an even function 
of $M$. (This can be seen by changing 
$p_0$ to $-p_0$ together with the change $M\rightarrow -M$ in (\ref{eq:33}),
in (\ref{eq:34}) and in
\be
M{^\prime_{\xi,B}}^2=-\xi e^2 M^2\int\frac{d^4p}{(2\pi)^3}
             (p_0 M + p^2)f_B(p)\frac{d\delta(p^2)}{dp^2}
              {\rm Re}\frac{1}{p^2+2p_0M+M^2+i\epsilon},
\ee
the last equation having been obtained by putting (\ref{eq:simple})
in (\ref{eq:Tmass}).) These considerations,
plus dimensional analysis, tell us that at high $T$, $M{^\prime_{\rm FG}}^2$ goes like
$T^2$, $M{^\prime_{\xi,F}}^2$ like $M^2(\ln|T/M|)^{n_1}$ 
and $M{^\prime_{\xi,B}}^2$ like $M^2(\ln|T/M|)^{n_2}$ 
($n_1$,$n_2$ being positive integers).  Since after $T^2$, the next
allowed term in $M{^\prime_{\rm FG}}^2$ is $M^2(\ln|T/M|)^{n_3}$ ($n_3$ a
positive integer), the general expression for ${M^\prime}^2$ at high $T$ can be written as
\be
{M^\prime}^2=e^2\Big[c_0 T^2 +c_3 M^2 \Big(\ln\frac{T}{M}\Big)^{n_3} 
     +\xi M^2 \Big(c_1\Big(\ln\frac{T}{M}\Big)^{n_1}
     +c_2\Big(\ln\frac{T}{M}\Big)^{n_2}\Big)\Big]
          \lb{eq:39}
\ee
where $c_0$,$c_1$,$c_2$,$c_3$ are constants, and we have dropped the modulus of the
 arguments of the logarithms since $T/M$ is positive. Terms independent
of $T$ have been neglected on the RHS of (\ref{eq:39}).

In principle there could be further constraints ruling out some term(s)
in (\ref{eq:39}), but a detailed calculation, described in Appendix \ref{app:masseqn},
reveals that all these terms are indeed present and that $n_1=1=n_2=n_3$.
Thus we actually have the high temperature equation
\be
{M^\prime}^2=e^2(\frac{ T^2}{8} +\frac{M^2}{8\pi^2}\ln\frac{T}{M}
     +\xi\frac{ M^2}{8\pi^2}\ln\frac{T}{M}).              \lb{eq:40}
\ee
The detailed calculation may also be viewed as a check on the argument
involving the degree of UV divergence mentioned before.

Using (\ref{eq:azero}) and (\ref{eq:40}) in (\ref{eq:totalmass}), we then have
\be
M^2=e^2\Bigg[\frac{T^2}{8}+(1+\xi)\frac{M^2}{8\pi^2}\ln\frac{T}{\sigma}\Bigg].
                                     \lb{eq:newmass}
\ee
On the RHS of (\ref{eq:newmass}) (as also in (\ref{eq:40})), 
we have not given the term $\sim e^2 M^2$; all other terms contributing
to the RHS of (\ref{eq:newmass}) (and (\ref{eq:40})) vanish in the limit
of large $T/M$. It is interesting to note that in (\ref{eq:newmass}), the
$\ln(M/\sigma)$ term from $\Sigma^{T=0}$ and the
$\ln(T/M)$ term from $\Sigma^\prime$ have exactly combined to yield
just a $\ln(T/\sigma)$ term, because it has been observed that similar 
combination also takes place in the case of the gauge boson self-energy
in the Yang-Mills theory \cite{weldon2}. A discussion of similar behaviour
in the case of three-point function, and, in general, $N$-point function
of gauge boson in the Yang-Mills theory, is to be found in 
Ref. \cite{brandt}.

The first important observation from 
(\ref{eq:newmass}) is that $M^2=e^2T^2/8$ to leading order in $T$, which
is a well-known result \cite{weldon}. Now, the gauge dependent part of
${\rm Re}\Sigma^\prime$ goes like $T$. 
Therefore in view of (\ref{eq:31}), one
would expect an $e^2\xi MT$ term in (\ref{eq:newmass}). The absence of such
a term shows that the part of ${\rm Re}\Sigma{^\prime_\xi}$ leading in
$T$ does not contribute to $M$.

To quantify the effect of this absence let us consider the equation
\be
M^2=e^2(\frac{T^2}{8}+ cMT)                   \lb{eq:41}
\ee
where $c$ is a constant. To leading order in $T$, the second term on
 the RHS is negligible, so that $M=O(eT)$. 
 This can now be used as an approximation in the second term, to give
\be
M^2=e^2\frac{T^2}{8}[1+O(e)]
\ee
showing that there is a correction of $O(e^2T)$ to $M$ (this conclusion
 can also be arrived at by solving (\ref{eq:41}) for $M$). So the absence
of the $e^2 MT$ term means that there is no such correction in any linear covariant gauge. It may be noted that the consequences of the
presence or absence of various {\it powers} of $T$ in (\ref{eq:newmass})
are to be taken seriously despite the presence of the scale $\sigma$, since
the $\sigma$ dependence is only logarithmic and is not multiplied with
 any power of $T$.

Finally, (\ref{eq:newmass}) shows that only the subleading $\ln T$ dependence of
${\rm Re}\Sigma{^\prime_\xi}$ contributes to $M$. Using the leading order
result $M=O(eT)$ as before to approximate the remaining terms on the RHS of (\ref{eq:newmass})
we infer that the $\xi$-dependence in (\ref{eq:newmass}) is $O(e^4 T^2)$,
apart from the logarithm.
As one certainly expects $O(e^4T^2)$ contribution to (\ref{eq:newmass}) from 
two loops, it is possible that the $\xi$-dependence which we have
 obtained will be canceled by the two-loop contribution. But it seems that this
issue can be decided only by an actual two-loop calculation. It is however clear
that an $e^2\xi MT$ term, being of $O(e^3T^2)$, was less likely to be canceled
by two-loop contribution. This shows the significance of the absence of such a term. 

We end this section by clarifying the following issue. Equation (\ref{eq:newmass})
has been obtained from the relation (\ref{eq:31}) which again is derived from
the usual dispersion relation (\ref{eq:Y}) (see Appendix \ref{app:massformula}).
In view of the discussion in the first paragraph of this section, one may wonder 
whether the terms correcting (\ref{eq:Y}), as given in (\ref{eq:ZZ}), can be
neglected even at $k=0$. In Appendix \ref{app:corrections}, we show that the
correction caused by these terms to (\ref{eq:newmass}) is indeed negligible.

\section{Conclusions}
        \lb{sec:conclusions}

We have carried out an investigation of gauge independence of the high-temperature
electron dispersion relation at one loop. While this independence is well-known
to leading order in $T$ by virtue of the gauge independence of the effective self-energy
at $O(T^2)$, our work takes the subleading 
$T$-dependence into account.
The analysis has been confined to linear covariant gauges and to the $k\gg eT$
and $k=0$ limits.

We have obtained an equation which governs the gauge dependence of the one-loop 
dispersion relation, stressed its analogy with the corresponding equation in the
gluon case, and consequently pointed out the possibility of a generalization of our
equation to all orders. From this equation obtained by us, the gauge independence
in the $k\gg eT$ limit follows in a straightforward way.

We have then shown that the effective mass is not affected by the leading 
gauge-dependent part of the
effective self-energy (going like $T$) and hence does not receive $O(e^2 T)$
correction in any gauge. While the effective mass is found to be influenced
by the subleading gauge-dependent part of the self-energy (going like $\ln T$), it is
possible that this will be canceled by two-loop contribution.

\subsection*{Acknowledgment}

The author thanks Prof. P. B. Pal for useful discussions, and also for going
through the manuscript and suggesting improvements.

\subsection*{Appendices}
\appendix
\section{Calculation of One-Loop Effective Electron Self-Energy}
                 \lb{app:calculation}

In this appendix, we first present some of the steps leading to 
(\ref{eq:azero}). Dimensional regularization of $\Sigma{^{T=0}_{\rm FG}}$
in $4-\epsilon^\prime$ dimensions gives 
\be
a{^{T=0}_{\rm FG}}=\frac{e^2}{(4\pi)^2}\Bigg[\frac{2}{\epsilon^\prime}
               -\gamma-1+\ln(4\pi)-2\int_0^1 dx\:x \ln\Big(x(1-x)K^2\Big)
               +O(\epsilon^\prime)\Bigg].
\ee
On the other hand, one easily obtains
\be
\Sigma{^{T=0}_{\xi}}=
-i\xi e^2\int\frac{d^4p}{(2\pi)^4}\frac{1}{(p^2+i\epsilon)^2}
      \Bigg[\rlap/p - \frac{p^2 \rlap/K +K^2 \rlap/p}
      {(p+K)^2+i\epsilon}\Bigg].              
\ee
The odd $\rlap/p$ term vanishes on integration. Evaluation of the rest 
leads to
\be
a{^{T=0}_{\xi}}=\frac{\xi e^2}{(4\pi)^2}\Bigg[\frac{2}{\epsilon^\prime}
               -\gamma-1+\ln(4\pi)-\int_0^1 dx\: \ln\Big(x(1-x)K^2\Big)
               +O(\epsilon^\prime)\Bigg].
\ee
Finally adding the counter-term to $a^{T=0}$, as fixed by the 
renormalization condition
\be
a{^{T=0}_{ren}}(K^2=\sigma^2)=0,
\ee
we arrive at (\ref{eq:azero}).

Next we present the details of the manipulations leading
to the equations (\ref{eq:16})-(\ref{eq:23}). The manipulations follow
Ref. \cite{weldon}, except for part of the derivation of
(\ref{eq:23}).

To obtain ${\rm Re}\Sigma{^\prime_{\rm FG}}$, one has to put (\ref{eq:8})
and (\ref{eq:11}) in the expression (\ref{eq:7}) for the self-energy, and
consider the relevant terms. Now change $p$ to $-p-K$ in the $f_F$-containing
term for convenience. Finally setting $p^2=0$, as allowed by $\delta(p^2)$,
yields (\ref{eq:16}).

Putting (\ref{eq:8}) and (\ref{eq:12}) in (\ref{eq:7}), one arrives at
${\rm Re}\Sigma{^\prime_\xi}$. It contains an $f_F$-containing part and
an $f_B$-containing part, as indicated in (\ref{eq:17}). 
In ${\rm Re}\Sigma{^\prime_{\xi,F}}$, changing $p$ to $-p-K$ and then setting
$p^2=0$ (allowed by the delta function) gives us (\ref{eq:21}).  
In ${\rm Re}\Sigma{^\prime_{\xi,B}}$, simplification leads to
\be
{\rm Re}\Sigma{^\prime_{\xi,B}}=\xi e^2\int\frac{d^4p}{(2\pi)^3}f_B(p)
                  \frac{d\delta(p^2)}{dp^2}
      \Bigg[\rlap/p - (p^2 \rlap/K +K^2 \rlap/p){\rm Re}\frac{1}
      {(p+K)^2+i\epsilon}\Bigg].               \lb{eq:simple}
\ee
The $\rlap/p$ term, being odd, drops out on integration. 
Now we make use of (\ref{eq:delta}), commute the 
integration over $p_0$ with the limit and the differentiation
involving $\lambda$ , and set $p^2=\lambda^2$
(allowed by the delta function). We then obtain
\bea
{\rm Re}\Sigma{^\prime_{\xi,B}} = \xi e^2 \int\frac{d^3\vec p}{(2\pi)^3}
      \lim_{\lambda\rightarrow 0} \frac{\partial}{\partial\lambda^2}\int dp_0
      f_B(p)\delta(p^2-\lambda^2)(K^2\rlap/p+\lambda^2\rlap/K) \nonumber \\
      {\rm Re}\frac{1}{K^2+2p.K+\lambda^2+i\epsilon}.
\eea
After integrating over $p_0$, the operations involving $\lambda$ are carried out.
While this last step is easily performed for the part proportional to $\lambda^2$
(since
\be
\lim_{\lambda\rightarrow 0}\frac{\partial}{\partial\lambda^2}
(\lambda^2 f(\lambda^2)) = f(0)                 \lb{trick}
\ee
with $f$ denoting a function with finite $\partial f/ \partial\lambda^2$
at $\lambda=0$ \cite{footnote3}), a more tedious algebra is to be worked out for the
remaining part, finally giving (\ref{eq:23}).

\section{Leading Order Behaviour of $\lowercase{a}$ and $\lowercase{b}$ at Zero Momentum}
       \lb{app:a-and-b} 
In this appendix, we work out the leading order behaviour of $a$ and $b$
at $k=0$ (the precise values were stated at the beginning of Sec.
\ref{sec:proof-for-mass}) {\it without performing a detailed calculation}.
Similar arguments will be used in Appendix \ref{app:corrections}.

At leading order $a$ and $b$ are $e^2 T^2$ times some function of $\omega$ and $k$
($T^2$ being deduced from UV power counting in ${\rm Re}\Sigma^\prime$).
Therefore to have the correct dimensions $a_{k=0}\sim e^2 T^2/{M_0}^2$
and $b_{k=0}\sim e^2 T^2/M_0$, where $M_0=\omega_{k=0}$ at leading order.
Putting these in the dispersion relation (\ref{eq:27a}) at $k=0$, namely,
\be
M_0+M_0 a_{k=0} +b_{k=0}=0,
\ee
gives us the well-known result \cite{weldon}
 $M_0\sim eT$. Using this, it follows that
$a_{k=0}$ is of $O(1)$ and $b_{k=0}$ is of $O(eT)$.

\section{Formula for the Effective Mass}
    \lb{app:massformula}

In this appendix, we prove the formula (\ref{eq:31}) for the effective mass.

The traces ${\rm Tr}(\rlap/K {\rm Re}\Sigma)$ and
${\rm Tr}(\rlap/u {\rm Re}\Sigma)$ are functions of
$\omega$ and $k$. The expressions for them are obtained from
(\ref{eq:sigmazero}), (\ref{eq:azero}), (\ref{eq:16}), 
(\ref{eq:21}) and (\ref{eq:23}). First of all we show that
these traces are both even functions of $k$. For the $\Sigma^{T=0}$ part,
 this is obvious from (\ref{eq:K^2}). For the $\Sigma^\prime$ part, we note that
in the rest-frame of the medium,
$k=|\vec K|$, and $\vec K$ can occur in the expressions for the traces only
through $\vec K.\vec p$ and $\vec K^2$. Odd power of $k$ can come from
$\vec K.\vec p=k|\vec p|{\rm cos}\theta$ where $\theta$ is the angle between
$\vec K$ and $\vec p$. This is also the only place where $\theta$ occurs
in the integrand. So changing $\theta$
to $\pi-\theta$ together with the change $k\rightarrow -k$, we establish
the even nature as a function of $k$.

For small $k$, therefore, we can write
\be
\frac{1}{4}{\rm Tr}(\rlap/K {\rm Re}\Sigma)
=h_0+h_1k^2+h_2k^4+\cdots
\ee
\be
\frac{1}{4}{\rm Tr}(\rlap/u {\rm Re}\Sigma)
=g_0+g_1k^2+g_2k^4+\cdots
\ee
where $h_i$ and $g_i$ are functions of $\omega$. Substitution of the above 
expressions into (\ref{eq:24}) and
 (\ref{eq:25}) gives
\be
a=\frac{1}{k^2}(h_0 -\omega g_0) +(h_1 -\omega g_1) +O(k^2),
                     \lb{eq:f1}
\ee
\be
b=-\frac{\omega}{k^2}(h_0 -\omega g_0) +
  (-\omega h_1 +\omega^2 g_1-g_0) +O(k^2).
                      \lb{eq:f2}
\ee
For the form factors $a$ and $b$ to remain analytic at $k=0$, the
relation 
\be
h_0=\omega g_0      \lb{eq:f3}
\ee
 must hold. We now put (\ref{eq:f1})
and (\ref{eq:f2}), subject to the constraint (\ref{eq:f3}), in
the dispersion relation (\ref{eq:27a}). Then we use the fact that for
any small $k$, the solution
\be
\omega=M+\omega_1 k+\omega_2 k^2+\cdots    \lb{eq:smallk}
\ee
where $M$ and $\omega_i$ are constants, must satisfy the dispersion  
relation. We also expand $h_i$ and $g_i$
in $k$ by first doing a Taylor expansion of them around
 $\omega=M$ as functions of $\omega$, and then putting
(\ref{eq:smallk}). Thus
\be
h_i(\omega)=h_i(M)+h{^\prime_i}(M)\omega_1 k+O(k^2)   \lb{eq:47}
\ee
with a similar equation for $g_i$. The equation that now results from 
the dispersion relation (\ref{eq:27a}) has a series in powers of $k$ 
alone on the LHS. Equating the coefficient of each power of $k$
separately to zero will determine the constants $M$ and $\omega_i$
of (\ref{eq:smallk}). Thus,
equating the constant term to zero gives
\be
M=g_0(M).         \lb{eq:50}
\ee
Multiplying both sides by $M$, we make use of
\be
h_0(M)=Mg_0(M)              \lb{eq:48}
\ee
(which follows from (\ref{eq:f3})) to finally arrive at
\be
M^2=h_0(M).                 \lb{eq:51}
\ee
This is the formula (\ref{eq:31}). One can alternatively use (\ref{eq:50})
as the formula for the effective mass and end up with the same results.

\section{Thermal Contribution to the Equation for the Effective Mass 
         at High Temperature}
                                          \lb{app:masseqn}

In this appendix we arrive at the equation (\ref{eq:40}), which expresses the 
high $T$
contribution to the equation for the effective
mass, by explicit calculation of the expressions (\ref{eq:33}),
(\ref{eq:34}), (\ref{eq:mb1}), (\ref{eq:mb2}) and (\ref{eq:mb3}).

We begin with $M{^\prime_{\rm FG}}^2$, given by (\ref{eq:33}). After the $p_0$
integration is carried out, the integrand is a function of $|\vec p|$ alone. 
Then integrating over the angles trivially, we are left with a one-dimensional
 integral. On rearrangement and use of the identity 
\be
f_B(p)-f_F(p)=2f_B(2p)              \lb{eq:identity}
\ee
we get
\bea
M{^\prime_{\rm{FG}}}^2&=&\frac{e^2}{2\pi^2}\int_0^\infty dp\:p[f_B(p)+f_F(p)]
                           \nonumber   \\
 &&  -\frac{e^2 M^2}{4\pi^2}\int_0^\infty dp\:pf_B(p){\rm Re}
          \frac{1}{p^2-M^2-i\epsilon}     \lb{eq:52}
\eea
Here and elsewhere in this appendix, $f_F(p)$ and $f_B(p)$ denote (\ref{eq:9})
and (\ref{eq:13}) with $|p.u|$ replaced by $p$. Now the first part of
(\ref{eq:52}) can be evaluated exactly by using
\be
\int_0^\infty dp\:p f_B(p)=\frac{\pi^2 T^2}{6},
\ee
\be
\int_0^\infty dp\:p f_F(p)=\frac{\pi^2 T^2}{12}.
\ee
The second part of (\ref{eq:52}) involves the integral
\bea
I&\equiv&\int_0^\infty dp\:p\frac{1}{e^{p/T}-1}{\rm Pr}\frac{1}{p^2-M^2 }
                                          \lb{eq:55} \\
 &=&\int_0^\infty dx\:x\frac{1}{e^x-1}{\rm Pr}\frac{1}{x^2-(M/T)^2 }.
                                           \lb{eq:Integral}
\eea
Here Pr denotes the principal value. The functional
dependence of $I$ on $T/M$  was determined on a computer
 for large $T/M$, yielding
\be
I\approx -\frac{1}{2}\ln\frac{T}{M}.      \lb{eq:55a}
\ee
Note that this logarithmic behaviour is in accord with the UV behaviour
 of (\ref{eq:55}) without the distribution function. Thus at high T
\be
M{^\prime_{\rm FG}}^2=e^2\Big(\frac{T^2}{8}+\frac{M^2}{8\pi^2}\ln\frac{T}{M}\Big)
                                   \lb{eq:56}
\ee
neglecting terms independent of $T$.

Next we turn to $M{^\prime_{\xi,F}}^2$, given by (\ref{eq:34}). Integrating
over $p_0$ and the angles, we are led to
\bea
M{^\prime_{\xi,F}}^2&=&\frac{\xi e^2 M^2}{8\pi^2}\int_0^\infty dp\:p
      f_F(p){\rm Re}\frac{1}{p^2-\frac{M^2}{4}-\frac{i\epsilon}{2}} \nonumber  \\
      &&   +\frac{\xi e^2 M^4}{32\pi^2} \int_0^\infty dp\:p
      f_F(p){\rm Re}\frac{1}{(p^2-\frac{M^2}{4}-\frac{i\epsilon}{2})^2}.    
                                  \lb{eq:57}
\eea
Moving on to $M{^\prime_{\xi,B}}^2$, integration over the angles in (\ref{eq:mb1}), 
(\ref{eq:mb2}) and (\ref{eq:mb3}) lead to
\be
M{^\prime_{\xi,B({\rm I})}}^2=-\frac{\xi e^2 M^2}{8\pi^2}\int_0^\infty dp\:p
      f_B(p){\rm Re}\frac{1}{p^2-\frac{M^2}{4}-\frac{i\epsilon}{2}}, 
\ee
\be
M{^\prime_{\xi,B({\rm II})}}^2=-\frac{\xi e^2 M^4}{32\pi^2} \int_0^\infty dp\:p
      f_B(p){\rm Re}\frac{1}{(p^2-\frac{M^2}{4}-\frac{i\epsilon}{2})^2},    
\ee
\be
M{^\prime_{\xi,B({\rm III})}}^2=-\frac{\xi e^2 M^2\beta}{8\pi^2}\int_0^\infty dp\:
          p^2 \frac{e^{\beta p}}{(e^{\beta p}-1)^2}
          {\rm Re}\frac{1}{p^2-\frac{M^2}{4}-\frac{i\epsilon}{2}}. 
\ee
Using the identity (\ref{eq:identity}) we then arrive at
\bea
M{^\prime_{\xi,F}}^2&+&M{^\prime_{\xi,B({\rm I})}}^2+ 
                       M{^\prime_{\xi,B({\rm II})}}^2
                    \nonumber    \\
               &=&-\frac{\xi e^2 M^2}{4\pi^2}\int_0^\infty dp\:p
               f_B(p){\rm Re}\frac{1}{p^2-M^2-2i\epsilon} \nonumber  \\
               &&   -\frac{\xi e^2 M^4}{4\pi^2} \int_0^\infty dp\:p
               f_B(p){\rm Re}\frac{1}{(p^2-M^2-2i\epsilon)^2}. 
                    \lb{eq:1111}   
\eea
While the first part again involves (\ref{eq:55}), the second part, being
ill-defined, is regularized in a way similar to (\ref{eq:reg}):
\bea
&&\int_0^\infty dp\:p f_B(p){\rm Re}\frac{1}{(p^2-M^2-2i\epsilon)^2}
                                   \nonumber \\
& =&\lim_{\lambda\rightarrow M}\frac{\partial}{\partial\lambda^2}
\int_0^\infty dp\:p f_B(p){\rm Re}\frac{1}{p^2-\lambda^2-2i\epsilon}.
                                \lb{eq:58}
\eea
The integral on the RHS of (\ref{eq:58}) is similar to $I$ in (\ref{eq:55})
and equals $-\frac{1}{2}\ln (T/\lambda)$ for high $T$. So the second part of
(\ref{eq:1111}) equals $-\xi e^2 M^2/16\pi^2$ which can be neglected as it
is independent of $T$. Thus we are left with the first part of
 (\ref{eq:1111}), so that
\be
M{^\prime_{\xi,F}}^2+M{^\prime_{\xi,B({\rm I})}}^2+ 
                       M{^\prime_{\xi,B({\rm II})}}^2
=\frac{\xi e^2 M^2}{8\pi^2}\ln\frac{T}{M}.  \lb{eq:59}
\ee
In $M{^\prime_{\xi,B({\rm III})}}^2$, we encounter the integral
\be
J\equiv\frac{1}{T}\int_0^\infty dp\:p^2
\frac{e^{p/T}}{(e^{p/T}-1)^2}
{\rm Pr}\frac{1}{p^2-{M_1}^2}   \lb{eq:62}
\ee
where $M_1=M/2$. Converting $J$ into a function of $T/M_1$ alone
(as we did in the case of $I$ in (\ref{eq:Integral})),
the functional dependence was found out for large $T/M_1$ using a
 computer, which gave $J\approx -\frac{1}{2}$.
  This constancy again agrees
with the expectation from the argument involving the UV behaviour.
Alternatively one can note that
\be
J=T\frac{\partial}{\partial T}\int_0^\infty dp\:p\frac{1}{e^{p/T}-1}
      {\rm Pr}\frac{1}{p^2-{M_1}^2}.   
\ee
Then using our knowledge of $I$ (see (\ref{eq:55}) and {\ref{eq:55a})),
we obtain $J\approx -\frac{1}{2}$ at large $T$.
Thus $M{^\prime_{\xi,B({\rm III})}}^2$ equals $\xi e^2 M^2/16\pi^2$ 
and can be neglected.
   
Therefore in the high-$T$ limit, ${M^\prime}^2$ receives contribution from (\ref{eq:56})
and (\ref{eq:59}), thereby leading us to (\ref{eq:40}).

\section{Terms Correcting the Usual Dispersion Relation : Estimation at 
        $\lowercase{k}=0$}
         \lb{app:corrections}

In this appendix, we estimate the correction to the equation (\ref{eq:newmass})
for the effective mass due to the terms in (\ref{eq:ZZ}) correcting 
the usual dispersion relation (\ref{eq:Y}).

Since $\omega_R=M$ at $k=0$, the dispersion relation (\ref{eq:ZZ})
at $k=0$ is given by
\be
f_R(\omega=M,k=0)+\frac{\gamma_0}{2}\frac{\partial}{\partial M}
f_I(M,0)-\frac{{\gamma_0}^2}{8}\frac{\partial^2}{\partial M^2}
f_R(M,0)=0,                \lb{eq:ZZ0}
\ee
where, from (\ref{eq:Z}),
\be
\gamma_0\equiv\gamma(k=0)=2\Bigg[\frac{\partial}{\partial M} f_R(M,0)\Bigg]^{-1}
                             f_I(M,0).       \lb{eq:gamma0}
\ee
Now from (\ref{eq:freal}) and (\ref{eq:fim}),
\be
f_R(M,0)=M+Ma_R(M,0)+b_R(M,0),
\ee
\be
f_I(M,0)=Ma_I(M,0)+b_I(M,0).
\ee
To leading order in $T$, $Ma_R(M,0)$ and $b_R(M,0)$ are $e^2T^2/M$ apart from
constant factors, as was explained in Appendix \ref{app:a-and-b} (we
shall put $M=M_0$ when we estimate the expressions we are interested 
in). Next there are a number of possible subleading terms. But we have 
already worked them out since (\ref{eq:newmass}) is just the equation $f_R(M,0)
=0$ multiplied by $M$ (see below (\ref{eq:50})). Thus
\be
f_R(M,0)=M-\frac{e^2 T^2}{8M}
     -e^2(1+\xi)\frac{ M}{8\pi^2}\ln\frac{T}{\sigma}.           \lb{eq:freal1}
\ee
(We would have got just this expression equated to zero in place of 
(\ref{eq:newmass}) if we used (\ref{eq:50}) in place of (\ref{eq:51}).)

At high temperature ${\rm Im}\Sigma$ goes like $T$. One way of seeing
this, which suffices for our purpose, is as follows. ${\rm Im}\Sigma$
is given by \cite{landsman,damping}
\be
{\rm Im}\Sigma(K)=[1-2f_F(K)]^{-1} {\rm Im}\Sigma_{11}(K).
\ee
For $T\gg \omega$, $T\gg k$ and $\omega^2-k^2>0$ it can be argued in the 
Feynman gauge that ${\rm Im}\Sigma{^\prime_{11}}$ is of order $\omega$
or $k$, since the integration involved is over two-body phase space 
\cite{footnote4}. Therefore in such a case ${\rm Im}\Sigma_{11}$
is independent of $T$ at leading order. Also for $\omega>0$ and $T\gg \omega$,
$[1-2f_F(K)]^{-1}\approx 2T/\omega$. Clearly, then, for our case
($\omega=M$, $k=0$, $T\gg M$), ${\rm Im}\Sigma$ goes like $T$.
We assume this leading order behaviour to hold in a general gauge, 
from considerations of gauge independence.  

It follows that to leading order in $T$, $Ma_I(M,0)$ and $b_I(M,0)$ are
$e^2 T$ apart from constant factors (no extra factor of $M$ required 
to match dimensions). Hence
\be
f_I(M,0)=d_1 e^2 T+d_2 e^2 M \Big(\ln\frac{T}{M}\Big)^{l}
                       \lb{eq:fim1}
\ee
where $d_1$, $d_2$ are constants and $l$ is a positive integer
(there is no $\sigma$, as ${\rm Im}\Sigma^{T=0}$ is finite and not
renormalized).
Terms independent of $T$ have been dropped in (\ref{eq:fim1}),
just as in (\ref{eq:freal1}). From the considerations of the previous paragraph,
 it easily follows that a subleading $\ln(T/M)$ term cannot in fact arise in 
${\rm Im}\Sigma$ in the Feynman gauge, but one cannot rule it out in a 
general gauge.

Now we shall put $M=M_0=eT/\sqrt 8$, the leading order value. Using 
(\ref{eq:freal1}) and (\ref{eq:fim1}), we then get 
\be
\gamma_0=d_1 e^2 T +\cdots
\ee
from (\ref{eq:gamma0}), and also
\be
\frac{\partial}{\partial M}f_I(M,0)=d_2 e^2\Big(\ln\frac{1}{e}\Big)^{l}
                               +\cdots,
\ee
\be
\frac{\partial^2}{\partial M^2}f_R(M,0)=-\frac{4\sqrt 2}{eT}+\cdots,
\ee
where the ellipsis in each equation indicates terms negligible in comparison
for $e\rightarrow 0$. 
Finally estimating the correction terms in (\ref{eq:ZZ0}), we find that
on the LHS of (\ref{eq:ZZ0}), the second term $\sim e^4(\ln(1/e))^l T$
while the third term $\sim e^3 T$. Since the $T$-independent terms 
dropped in (\ref{eq:freal1}) are $\sim e^2 M \sim e^3 T$, we are clearly 
justified in neglecting the second and third terms on the LHS of (\ref
{eq:ZZ0}), as we did in Sec. \ref{sec:proof-for-mass}.


\begin {thebibliography}{}
\bibitem{symbol}
     We reserve the symbol $M$ for the effective electron mass including the
     subleading $T$-dependence.
\bibitem{footnote1}
     As will be pointed out in the next section, this does not happen for
     $k\ll eT$.
\bibitem{similar}
     Similar idea has been used for neutrino dispersion relation in
     Ref. \cite{nieves}.
\bibitem{footnote4}
     See Appendix A of Ref. \cite{weldon}. Similar argument for gauge 
     boson self-energy is to be found in Sec. IIIA of the paper cited in
     Ref. \cite{weldon2}.
\end{thebibliography}
\end{document}